\documentclass[aas2pp4]{aastex}
\usepackage{lscape}

\slugcomment{}

\ifdim\textheight>25.2cm\textheight=25.0cm\fi

\shorttitle{AGNs and the Truncation of Star Formation}
\shortauthors{Brown {\it et al.}}

\begin{document}

\title{Active Galactic Nuclei and the Truncation of Star Formation in K+A Galaxies}

\author{
Michael J. I. Brown\altaffilmark{1},
John Moustakas\altaffilmark{2,3},
Nelson Caldwell\altaffilmark{4},
David Palamara\altaffilmark{1},
Richard J. Cool\altaffilmark{5,6,7}, 
Arjun Dey\altaffilmark{8},
Ryan Hickox\altaffilmark{9},
Buell T. Jannuzi\altaffilmark{8},
Stephen S. Murray\altaffilmark{9},
Dennis Zaritsky\altaffilmark{5}
}
\altaffiltext{1}{School of Physics, Monash University, Clayton, Victoria 3800, Australia}
\altaffiltext{2}{Center for Astrophysics and Space Sciences, University of California, San Diego, 9500 Gilman Drive, La Jolla, California, 92093-0424}
\altaffiltext{3}{Center for Cosmology and Particle Physics, 4 Washington Place, New York University, New York, NY 10003}
\altaffiltext{4}{Smithsonian Astrophysical Observatory, 60 Garden Street, Cambridge, MA 02138}
\altaffiltext{5}{Steward Observatory, University of Arizona, 933 North Cherry Avenue, Tucson, AZ 85721}
\altaffiltext{6}{Princeton University Observatory, Peyton Hall, Ivy Lane, Princeton, NJ 08544}
\altaffiltext{7}{Hubble Fellow}
\altaffiltext{8}{National Optical Astronomy Observatory, Tucson, AZ 85726-6732}
\altaffiltext{9}{Harvard-Smithsonian Center for Astrophysics, Cambridge, MA 02138}
\email{Michael.Brown@sci.monash.edu.au}

\begin{abstract}
We have searched for active galactic nuclei (AGNs) in K+A galaxies, using multiwavelength 
imaging and spectroscopy in the Bo\"otes field of the NOAO Deep Wide-Field Survey.
The K+A galaxies, which have had their star formation rapidly truncated, are 
selected via their strong Balmer absorption lines and weak $H\alpha$ emission.
Our sample consists of 24 K+A galaxies selected from 6594 $0.10<z<0.35$ galaxies brighter than 
$I=20$ with optical spectroscopy from the AGN and Galaxy Evolution Survey. Two thirds of the 
K+A galaxies are likely ongoing galaxy mergers, with nearby companion galaxies or tidal tails.
Galaxy mergers may be responsible for the truncation of star formation,
or we are observing the aftermath of merger triggered starbursts. As expected,
the optical colors of K+A galaxies largely fall between blue galaxies with ongoing star formation 
and red passive galaxies. However, only 1\% of the galaxies with colors between the red
and blue populations are K+A galaxies, and we conclude that the truncation of star formation
in K+A galaxies must have been unusually abrupt ($\lesssim 100 ~{\rm Myr}$).
We examined the AGN content of K+A galaxies with both optical emission-line ratios 
(BPT diagrams) and {\it Chandra} X-ray imaging.  At least half of 
all K+A galaxies display the optical emission-line ratios of AGNs, and a third
of $M_R<-22$ K+A galaxies host AGNs with X-ray luminosities of $\sim 10^{42}~{\rm erg~s^{-1}}$.
The faintest K+A galaxies do not show clear evidence for hosting AGNs, having
emission-line ratios consistent with photoionization by massive stars and few X-ray detections.
We speculate that two mechanisms may be responsible for the truncation of star 
formation in K+A galaxies, with AGN feedback only playing a role in  $M_R\lesssim -20.5$ galaxies.
\end{abstract} 

\keywords{galaxies: active -- galaxies: evolution -- galaxies: individual (IRAS~14344+3451) -- galaxies: interactions -- X-rays: galaxies}

\section{INTRODUCTION}
\label{sec:intro}

The stellar mass contained within the red galaxy population has doubled over the 
past 7 Gyr \markcite{bel04,bro07,fab07}(e.g., {Bell} {et~al.} 2004; {Brown} {et~al.} 2007; {Faber} {et~al.} 2007). Almost all red galaxies
have low specific star formation rates \markcite{fuk04,hem08}(e.g., {Fukugita} {et~al.} 2004; {Helmboldt}, {Walterbos}, \&  {Goto} 2008), so this rapid 
stellar mass growth cannot result from in-situ star formation alone. 
Mergers of red galaxies can redistribute, but not significantly increase, the stellar mass within the 
red galaxy population. For the stellar mass to rapidly increase within the red galaxy population, 
stellar mass must be transferred from the blue galaxy population to the red galaxy population \markcite{bel04}({Bell} {et~al.} 2004). 
For this to happen, star formation must be truncated within blue galaxies.

Galaxies that have had their star formation abruptly truncated within the
past several hundred million years can be identified spectroscopically \markcite{dre83,bor84,cou87,zab96}(e.g., {Dressler} \& {Gunn} 1983; {Boroson} \& {Oke} 1984; {Couch} \& {Sharples} 1987; {Zabludoff} {et~al.} 1996).
These galaxies, which lack massive OB stars but retain longer lived A stars, 
have weak nebular emission-lines (relative to star forming galaxies) and 
a prominent Balmer absorption series. These galaxies are commonly known as post-starbursts \markcite{cou87}(e.g {Couch} \& {Sharples} 1987), 
E+A galaxies \markcite{dre83}(e.g {Dressler} \& {Gunn} 1983), or K+A galaxies \markcite{fra93}(e.g {Franx} 1993), as their stellar continuum 
can be approximated by combining K and A star spectra. 

There are several plausible mechanisms for the rapid truncation of star formation
in galaxies, including strangulation \markcite{lar80,bal00b}({Larson}, {Tinsley}, \& {Caldwell} 1980; {Balogh} \& {Morris} 2000), supernova feedback \markcite{efs00}(e.g., {Efstathiou} 2000), 
virial shock heating \markcite{dek06}({Dekel} \& {Birnboim} 2006) and feedback from active galactic nuclei \markcite{sil98,wyi03,bow06,cro06,hop06}(AGNs; e.g., {Silk} \& {Rees} 1998; {Wyithe} \& {Loeb} 2003; {Bower} {et~al.} 2006; {Croton} {et~al.} 2006; {Hopkins} {et~al.} 2006a). 
Models incorporating AGN feedback have achieved prominence within 
the past decade, as they can reproduce the observed properties of low
redshift galaxies \markcite{bow06,cro06,hop06b}(e.g., {Bower} {et~al.} 2006; {Croton} {et~al.} 2006; {Hopkins} {et~al.} 2006b) and there is strong observational evidence for 
AGN feedback in nearby galaxy clusters \markcite{fab03a,fab03b}(e.g., {Fabian} {et~al.} 2003a, 2003b). However, galaxy models 
with AGN feedback can have more free parameters than competing models and while AGN activity is 
relatively easy to detect, proving that it is (solely) responsible for truncating star 
formation is non-trivial.

AGN host galaxies have been studied for decades, and while the primary goal
of these studies is understanding what triggers AGN activity, 
such studies can also explore the connections between AGN activity and star formation.
Examples of quasars with K+A host galaxies are well known \markcite{bor84,brot99,brot04}({Boroson} \& {Oke} 1984; {Brotherton} {et~al.} 1999, 2004),
although there are also powerful AGNs hosted by starbursts \markcite{gal02}(e.g., {Gallagher} {et~al.} 2002) and 
AGNs hosted by galaxies with old stellar populations \markcite{nol01}(e.g., {Nolan} {et~al.} 2001). 
While the presence of powerful AGNs in star forming galaxies could be evidence for AGN 
feedback playing a negligible role in galaxy evolution, there are scenarios where AGN feedback 
only plays a significant role after millions of years of AGN activity \markcite{hop06}(e.g., {Hopkins} {et~al.} 2006a).

Recently there have been a number of studies of the AGN content of K+A galaxies, 
with the principal goal of determining if AGNs could plausibly truncate star formation.
These studies suggest a correlation between AGN activity and the rapid
truncation of star formation, although a causal link has yet to be established.
K+A galaxies selected from the Sloan Digital Sky Survey \markcite{yor00}(SDSS; {York} {et~al.} 2000) typically have emission-line ratios 
comparable to LINERs \markcite{hec80}(Low Ionization Nuclear Emission-line Regions; {Heckman} 1980) and 
Seyferts, consistent with AGN activity in these galaxies \markcite{yan06}({Yan} {et~al.} 2006). 
A quarter of K+A galaxies have blue cores, and these cores have spectra similar to LINERs \markcite{yang06}({Yang} {et~al.} 2006). 
\markcite{tre07}{Tremonti}, {Moustakas}, \&  {Diamond-Stanic} (2007) observed Mg~II~$\lambda\lambda2796,2803$ absorption lines with blueshifts of 
$\sim 1000~{\rm km~s^{-1}}$ in 10 of 14 $z\sim 0.6$ post-starburst galaxies, and hypothesized 
that these outflows were launched by AGNs that also truncated star formation within these galaxies. 
Stacking of {\it Chandra} images of $z\sim 0.8$ post-starburst galaxies reveals a 
population X-ray sources with a mean luminosity of $\sim 3\times 10^{41}~{\rm erg~s^{-1}}$ \markcite{geo08}({Georgakakis} {et~al.} 2008).
These studies suggest a link between AGN activity and K+A galaxies.

In this paper, we examine the AGN content of $0.10<z<0.35$ K+A galaxies using the optical 
emission-line diagnostic diagrams of \markcite{bal81}{Baldwin}, {Phillips}, \&  {Terlevich} (1981, BPT diagrams) and X-ray imaging with the {\it Chandra X-ray Observatory}.
Our principal goal is to measure the fraction of K+A galaxies that host AGNs, and 
thus determine if AGN feedback is a {\it plausible} mechanism for rapidly truncating star formation. 
We caution that even if all of the K+A galaxies in our sample host an AGN, this finding would be consistent
with, but not direct evidence for, a causal link between AGN activity and the truncation of star formation.
We also note that if none of the K+A galaxies in our sample host an AGN, one could develop 
models where the AGN activity is so brief that it would be rarely observed, although one may consider such models contrived.
Despite these limitations and caveats, the plausibility of AGN feedback models does depend
on the observed fraction of K+A galaxies that currently host AGNs.

The structure of this paper is as follows. In \S\ref{sec:sample} we describe the selection of our 
K+A galaxy sample from the AGN and Galaxy Evolution Survey of  Bo\"otes. We discuss
the NOAO Deep Wide-Field Survey optical imaging, photometry and rest-frame properties of the 
K+A galaxies in \S\ref{sec:optical}. The AGN content of K+A galaxies, inferred from optical emission 
line ratios and {\it Chandra} X-ray imaging is discussed in \S\ref{sec:bpt} and \S\ref{sec:chandra} respectively. 
We summarize our principal conclusions in \S\ref{sec:summary}. Throughout this paper we use Vega-based magnitudes and 
adopt a flat cosmology with $\Omega_m=0.25$ and $H_0=72 ~{\rm km~s^{-1}~Mpc^{-1}}$, which is consistent with the cosmological
parameters of \markcite{spe07}{Spergel} {et~al.} (2007). 

\section{K+A GALAXY SAMPLE SELECTION}
\label{sec:sample}

We selected our sample of K+A galaxies from the AGN and Galaxy Evolution Survey (AGES; C.~S.~Kochanek et~al. in~prep.),
a spectroscopic survey of $7.9~{\rm deg}^2$ of the Bo\"otes  field of the NOAO Deep Wide-Field Survey \markcite{jan99}(NDWFS; {Jannuzi} \& {Dey} 1999). 
AGES targeted all optically extended sources brighter than $I=18.5$ and $\gtrsim 20\%$ of extended sources with magnitudes of
$18.5<I<20.0$. AGES also targeted other galaxies, using a variety of selection criteria (e.g., infrared colors), 
but we only include these objects in our $K+A$ sample if they were also part of the main $I$-band selected galaxy sample.

The spectra were obtained with Hectospec, a 300-fiber robotic spectrograph with a $1^\circ$ field-of-view 
on the 6.5-m MMT telescope \markcite{fab98,rol98}({Fabricant} {et~al.} 1998; {Roll}, {Fabricant}, \& {McLeod} 1998). The wavelength range of the spectra is   
$\simeq 3700~{\rm \AA}$ to $\simeq 9200~{\rm \AA}$, and the instrumental resolution is 6~\AA.
Spectra were extracted and classified using two independent 
reduction pipelines and then verified by visual inspection.  One pipeline is based upon the SDSS spectroscopic
data reduction software while the other is a set of customized IRAF scripts.
The spectra were flux calibrated by assigning several Hectospec fibers to F stars \markcite{yor00}(identified from the SDSS; {York} {et~al.} 2000), 
and then comparing their observed spectra with \markcite{kur92}{Kurucz} (1992) stellar atmosphere models.
To reliably select K+A galaxies using accurate measurements of galaxy emission and 
absorption lines, we require a median continuum signal-to-noise of 4 per pixel.
We excluded spectra if they were badly contaminated by a red light leak from an LED 
in the fiber positioner or suffered from poor flux calibration due to problems with the 
operation of the atmospheric dispersion corrector (even when reliable redshifts were possible).

To select K+A galaxies, we use the equivalent width of the ${\rm H\alpha}$ emission-line and 
the \markcite{wor97}{Worthey} \& {Ottaviani} (1997) indices for the ${\rm H\delta}$ and ${\rm H\gamma}$ absorption lines. 
Our approach is similar to that used in much of the literature. K+A galaxies are
identified by searching for spectra with strong Balmer absorption and weak nebular 
emission (e.g., ${\rm [OII]}~\lambda 3727$, ${\rm H\beta}$, ${\rm H\alpha}$) relative to star forming galaxies \markcite{zab96,qui04,yan08}(e.g., {Zabludoff} {et~al.} 1996; {Quintero} {et~al.} 2004; {Yan} {et~al.} 2008). 
While the selection criteria used in the literature are based on the same underlying principle, they
vary in detail and select different samples of objects, and we return to this point later.
The strength of the ${\rm H\alpha}$ emission-line is a very strong function of the number 
UV photons produced by massive OB stars. If star formation is completely truncated in a 
galaxy after a period of star formation, the ${\rm H\alpha}$ emission-line 
will disappear from the integrated galaxy spectrum within $10~{\rm Myr}$. 
The ${\rm H\delta}$ and ${\rm H\gamma}$ absorption lines are particularly strong in A star spectra, so these 
absorption lines remain prominent in galaxy spectra for $\sim 300~{\rm Myr}$ after the truncation of star formation. 

Our principal selection criteria for K+A galaxies are
\begin{eqnarray}
{\rm (H\delta_A + H\gamma_A) / 2 } & > & 3 \\ \nonumber
{\rm log( H\alpha~ EW)} & < & 0.2 \times  (H\delta_A + H\gamma_A) / 2, 
\end{eqnarray}
where ${\rm H\delta_A}$ and ${\rm H\gamma_A}$ are the spectral indices of \markcite{wor97}{Worthey} \& {Ottaviani} (1997) 
and ${\rm H\alpha~ EW}$ is the equivalent width of the $H\alpha$ emission-line (in units of ${\rm \AA}$).
Using the stellar population synthesis models of \markcite{bc03}{Bruzual} \& {Charlot} (2003) with Solar metallicity, we find that a galaxy 
that has had its star formation completely truncated after $10~{\rm Gyr}$ of continuous 
star formation will meet our selection criteria for $300~{\rm Myr}$. 

We caution that ${\rm H\alpha}$ emission can also be produced by AGNs as well as star formation, 
and some K+A galaxies that host AGNs will be absent from our sample. For example, IRAS~14344+3451 (J2000 {\it R.A.}=14:36:31.99 
{\it Decl.}=+34:38:29.5) is a $z=0.349$ obscured quasar, hosted by a post-starburst galaxy undergoing a merger, 
that is excluded from our sample as it has strong ${\rm H\alpha}$ emission.
While one could attempt to add such objects to the sample by hand, this greatly complicates the selection function 
so we have stuck with a simple selection criterion that can be easily modeled.
Our estimates of the fraction of K+A galaxies that host an AGN are thus lower limits.

\begin{figure}[h]
\plotone{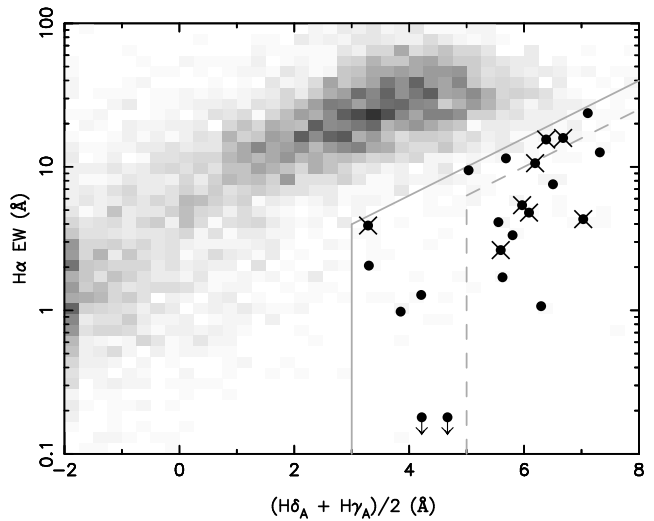}
\caption{${\rm H\alpha}$ emission-line equivalent width as a function of the \markcite{wor97}{Worthey} \& {Ottaviani} (1997) 
indices for the ${\rm H\delta}$ and ${\rm H\gamma}$ 
absorption lines. Our selection criterion for spectra with a signal-to-noise of 8 (4) or more per pixel is shown with the 
solid (dashed) line. Our final sample of K+A galaxies is shown with the black circles, with {\it 
Chandra} sources highlighted with crosses. For comparison, the greyscale shows the distribution of $I$-band selected 
AGES galaxies with redshifts of $0.10<z<0.35$ and magnitudes of $I<20$, including sources manually rejected from the post-starburst sample.
\label{fig:sel}}
\end{figure}

\begin{figure*}
\resizebox{7in}{!}{\includegraphics{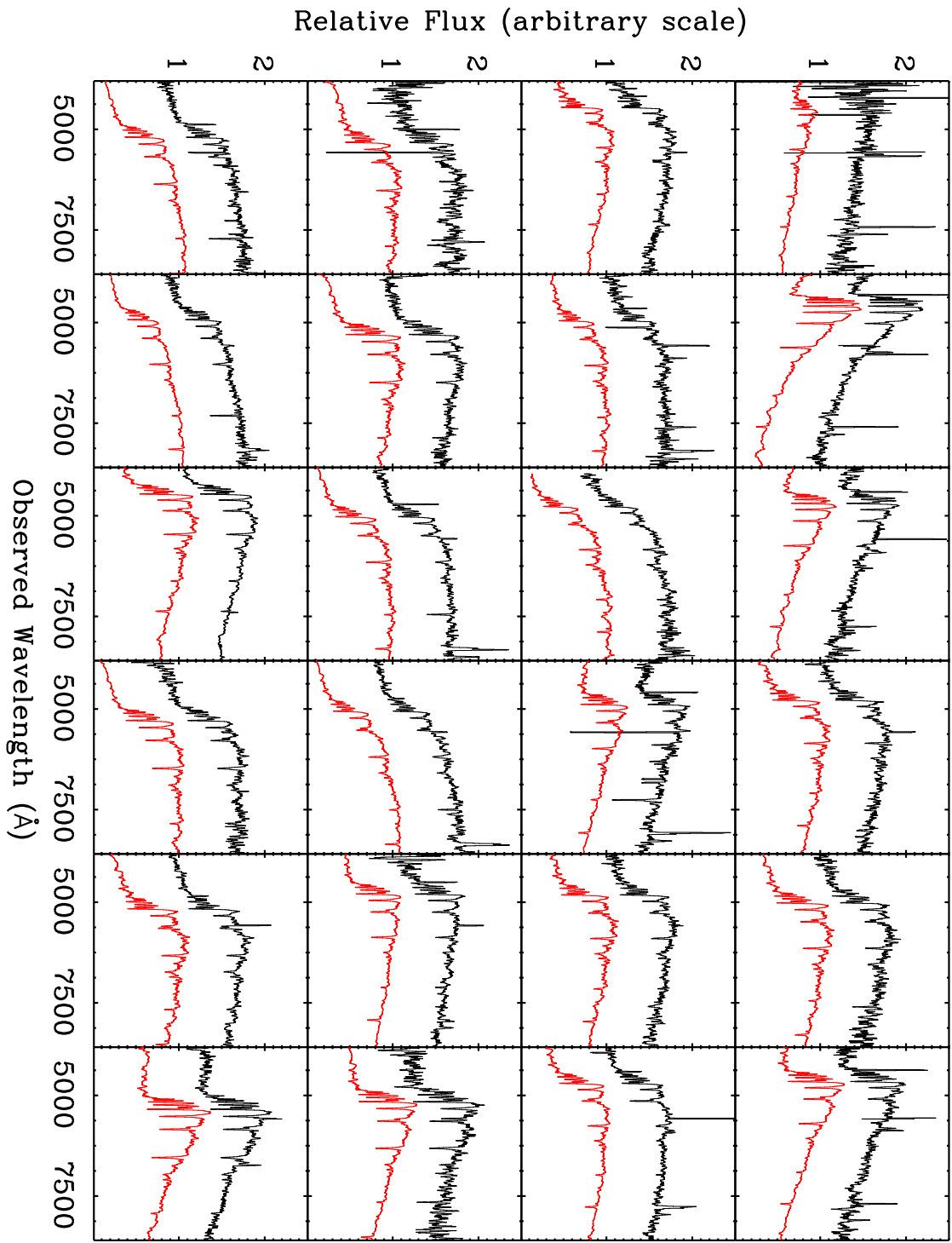}}
\caption{Spectra of the Bo\"otes K+A galaxies. The spectra are listed in order of absolute magnitude, with the top row
containing the faintest objects and the bottom row containing the brightest objects. In each panel we show the best-fit
continuum model (red line) underneath the AGES spectrum (black line). Although many of the galaxies have
nebular emission-lines, the equivalent widths of these lines are less than those of star forming galaxies with comparable 
Balmer absorption lines.
\label{fig:spectra}}
\end{figure*}

In addition to our principal selection criteria, we employ a number of other 
criteria and techniques to reduce contamination and improve the sample definition.
We limit our redshift range to $0.10<z<0.35$, so we have wavelength coverage of both the Balmer series and 
the emission-lines used in BPT diagrams.
The lower redshift limit also reduces (but does not eliminate) aperture bias, where a small and 
unrepresentative fraction of the galaxy flux enters the $1.5^{\prime\prime}$ Hectospec fiber.
Measurements of $H\delta_A$ and $H\gamma_A$ can be corrupted by residuals from the 5577~\AA~ sky line, 
so we measure the strength of the Balmer absorption using only one of these indices when 5577~\AA~ 
(potentially) contaminates the other index.
Measurement errors broaden the observed locus of star forming galaxies, so for spectra with a 
signal-to-noise per pixel between 4 and 8, we adopt the conservative selection criteria 
\begin{eqnarray}
{\rm (H\delta_A + H\gamma_A) / 2 } & > & 5 \\ \nonumber
{\rm log( H\alpha~ EW) } & < & -0.2 + 0.2 \times  (H\delta_A + H\gamma_A) / 2.
\end{eqnarray}
Contamination can be significant when selecting a very small subset of objects from a large population, 
so we visually inspected all of our K+A candidates and manually excluded 17 contaminants.
Manually rejected objects included spectra with night sky line residuals and spectra where a
red light leak had not been automatically flagged.
Our selection criteria, K+A galaxies and other $I<20$ galaxies are plotted in Figure~\ref{fig:sel}.
Of the 6954 $I$-band selected galaxies with $I<20$, $0.10<z<0.35$ and good AGES spectra, 24 were selected as K+A galaxies. 
The low percentage (0.3\%) of K+A galaxies in AGES is comparable to what has been observed in previous studies \markcite{zab96,qui04}(e.g., {Zabludoff} {et~al.} 1996; {Quintero} {et~al.} 2004), even though the selection criteria differ.
The coordinates and redshifts of the K+A galaxies are provided in  Table~\ref{table:summary} and we plot their spectra 
in  Figure~\ref{fig:spectra}.

\begin{figure*}
\plotone{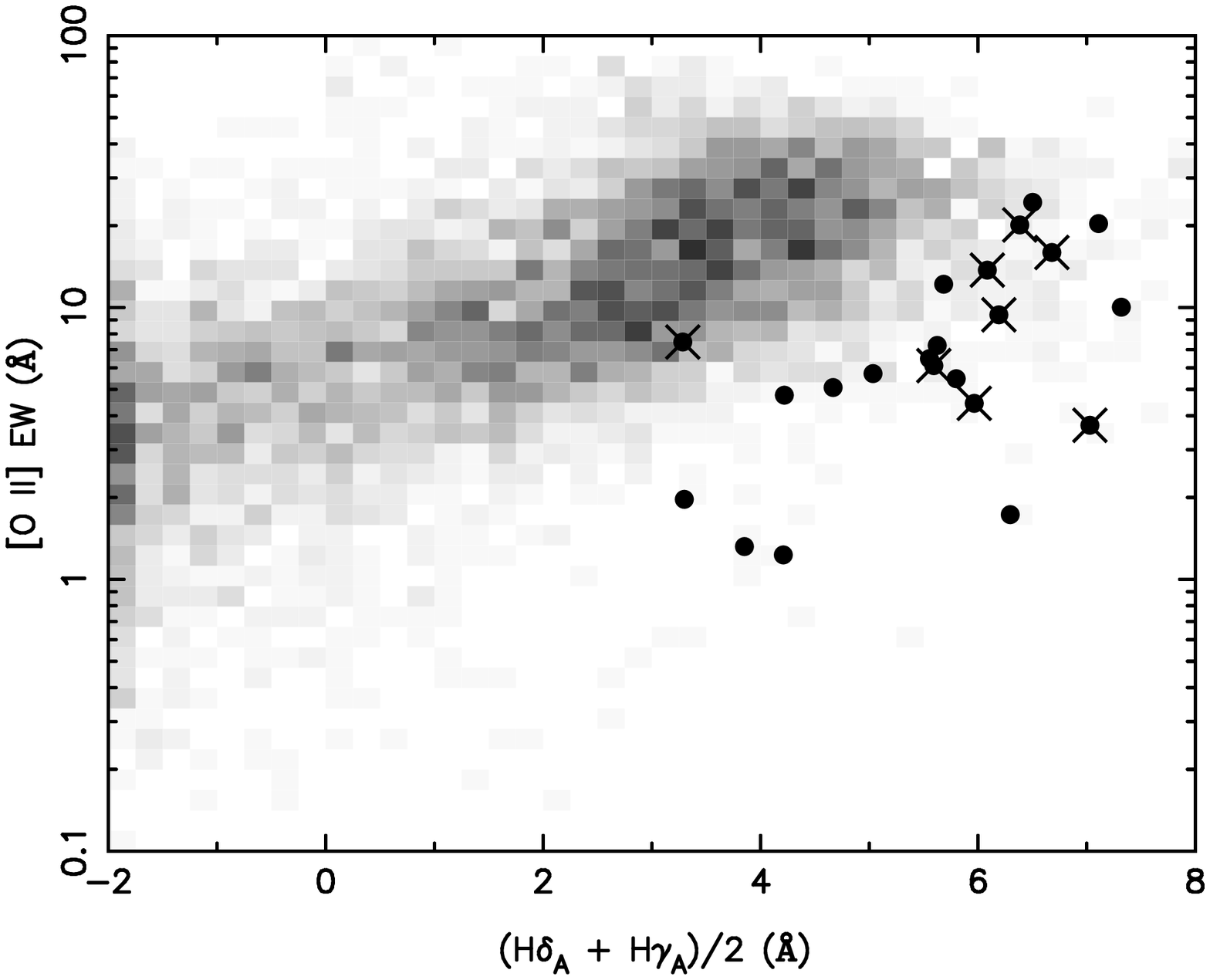}\plotone{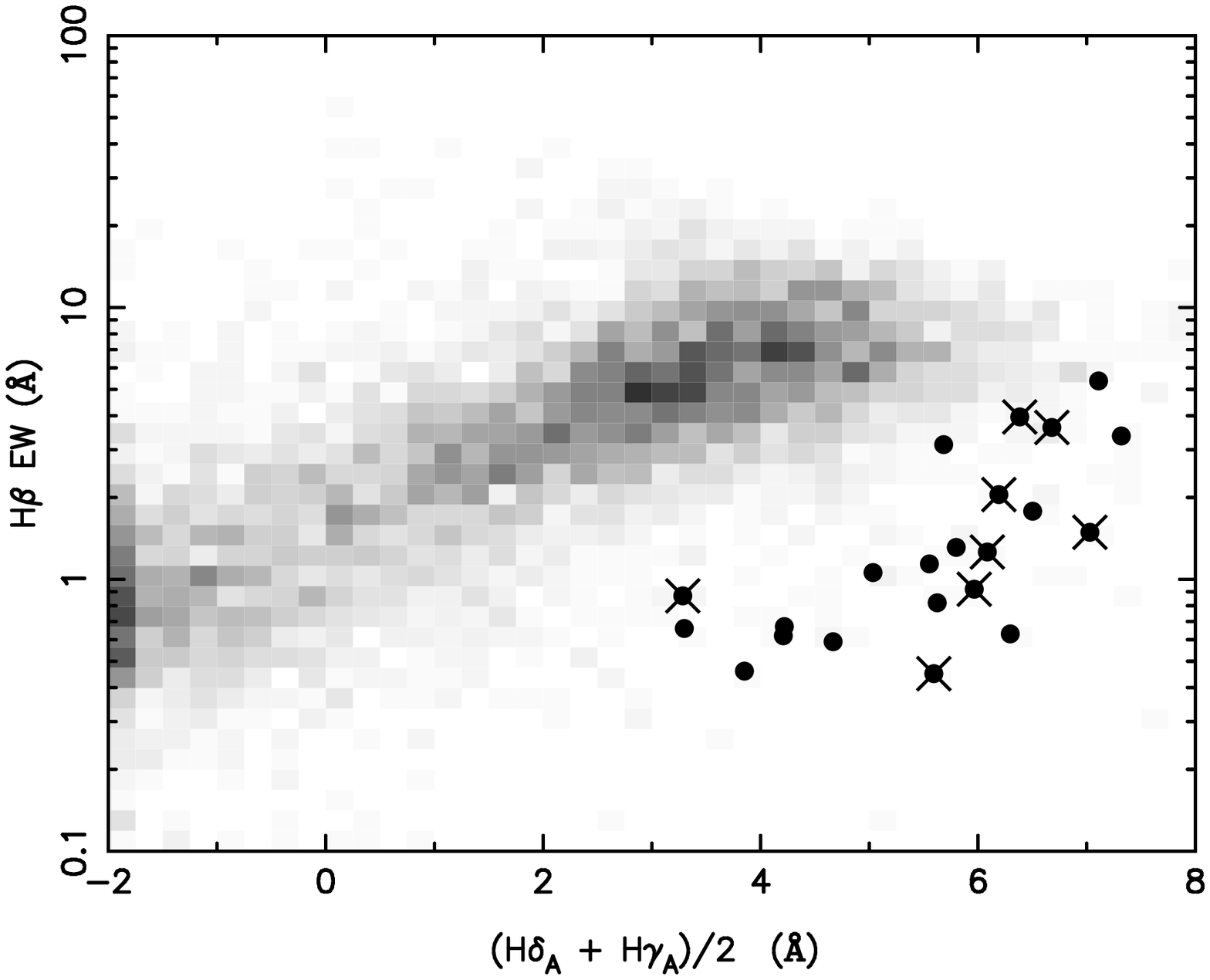}
\caption{${\rm [O~II]}$ and ${\rm H\beta}$ emission-line equivalent width as a function of 
the \markcite{wor97}{Worthey} \& {Ottaviani} (1997) indices for ${\rm H\delta}$ and ${\rm H\gamma}$.
Our final sample of K+A galaxies is shown with the black circles, with {\it Chandra} sources highlighted with crosses,
while the greyscale shows the distribution of other $I<20$ AGES galaxies with $0.10<z<0.35$.
Many of our K+A galaxies have weak ${\rm [O~II]}$ and ${\rm H\beta}$ relative to the locus of star-forming galaxies.
However, some of our K+A galaxies have relatively strong ${\rm [O~II]}$ or ${\rm H\beta}$ emission, and these objects would
not have satisfied some of the K+A selection criteria used in the literature \markcite{zab96}(e.g., {Zabludoff} {et~al.} 1996).
\label{fig:comp}}
\end{figure*}

A variety of K+A and post-starburst selection criteria are used in the literature. Consequently,
there is a genuine concern that conclusions derived from one sample will not be applicable
to other samples. To address (but not resolve) this concern, in 
Figure~\ref{fig:comp} we plot the equivalent width of the ${\rm [O~II]}$ and ${\rm H\beta}$ 
emission-lines as a function of the ${\rm H\delta_A}$ and ${\rm H\gamma_A}$ spectral indices.
Our K+A galaxy sample has weak ${\rm H\beta}$ relative to the star-forming locus, and at least $80\%$ of our objects
would satisfy the ${\rm H\beta}$ selection criterion of \markcite{qui04}{Quintero} {et~al.} (2004). Although our K+A galaxies have 
weaker ${\rm [O~II]}$ than galaxies in the star-forming locus, the difference is relatively small 
and $\sim 80\%$ of our K+A galaxies would be excluded from samples using the 2.5~\AA~and 5~\AA~${\rm [O~II]}$ 
emission-line equivalent width criteria of \markcite{zab96}{Zabludoff} {et~al.} (1996) and \markcite{tra04}{Tran} {et~al.} (2004) respectively.
As noted by \markcite{yan06}{Yan} {et~al.} (2006), ${\rm [O~II]}$ can result from AGN emission, and almost all of our K+A galaxies with 
X-ray counterparts (\S\ref{sec:chandra}) would be excluded from samples using an ${\rm [O~II]}$ selection criterion. 
Historically, ${\rm [O~II]}$ was used not because it was preferred for physical reasons, but because detectors were
more sensitive at those wavelengths. While our use of ${\rm H\alpha}$ makes some comparisons with previous 
work more difficult, it is a good choice if one is more concerned about detecting recent star formation.

\section{OPTICAL IMAGING AND PHOTOMETRY}
\label{sec:optical}

To determine if galaxy mergers are associated with K+A galaxies, 
we searched for evidence of ongoing mergers using NDWFS optical imaging.
As mergers are transitory, some galaxies that have undergone a merger within the 
past $\sim 300~{\rm Myr}$ may seem undisturbed in the NDWFS imaging.
As merger of two galaxies (initially separated by $\sim 30~{\rm kpc}$) can take up to a 
billion years  \markcite{tay01,boy08}(e.g., {Taylor} \& {Babul} 2001; {Boylan-Kolchin}, 
{Ma}, \&  {Quataert} 2008), merging K+A galaxies may be end products of merger triggered
starbursts. While such caveats should be kept in mind, if a higher fraction of K+A galaxies 
are undergoing mergers than the overall galaxy population, this would be evidence for a 
link (albeit indirect) between K+A galaxies and galaxy mergers.

\begin{figure*}
\resizebox{7in}{!}{\includegraphics{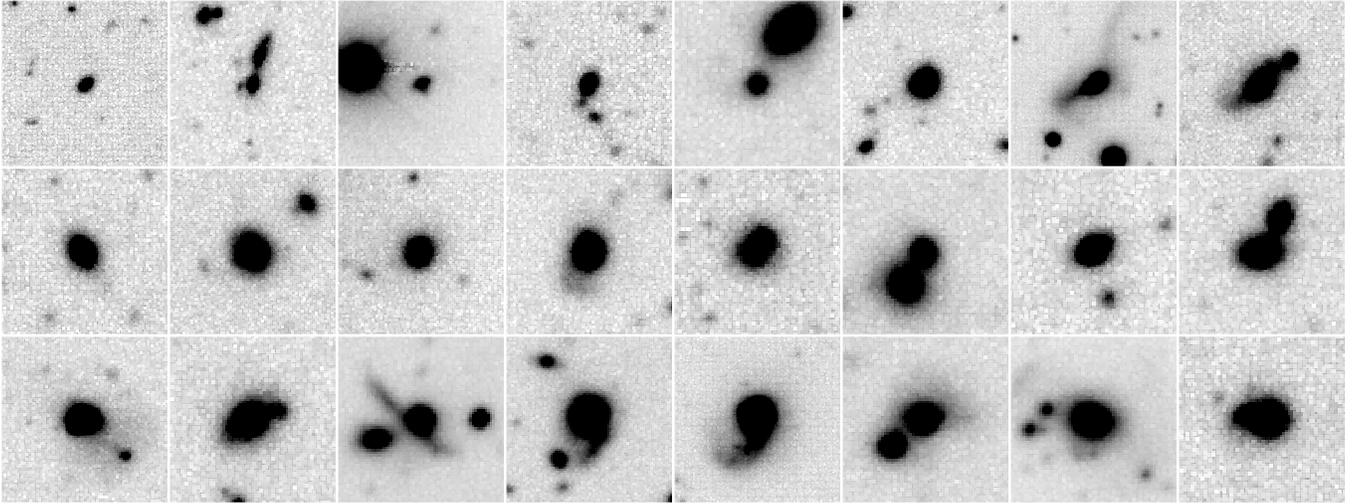}}
\caption{NDWFS $I$-band $20^{\prime\prime}$ field-of-view images of the K+A galaxies. 
As in Table~\ref{table:summary}, the objects are ordered by absolute magnitude, with 
faint objects in the top row and bright objects in the bottom row. We find that two thirds of the K+A galaxies show 
evidence for ongoing mergers, including nearby companion galaxies and tidal tails. All but one of the 
$M_R<-22$ K+A galaxies (bottom row) has a nearby companion galaxy or tidal tail. 
\label{fig:cutout}}
\end{figure*}

To determine if galaxy mergers play a role in the truncation of star formation, we searched
the NDWFS optical imaging for companion galaxies within a projected distance of $30~{\rm kpc}$ 
from each K+A galaxy. To reduce contamination of the companion sample, we used the 
photometric redshifts of \markcite{bro08}{Brown} {et~al.} (2008) to exclude companions whose photometric 
redshifts differed from the K+A galaxy spectroscopic redshifts by more than $0.2$. 
We also excluded companion galaxies fainter than 10\% of the K+A galaxy luminosity,
as very faint galaxies can have large photometric redshift errors.
An automated search of an object catalog
will invariably miss some companion galaxies (e.g., due to blending) but is less likely
to suffer from biases than visual searches. While only 17\% of the overall 
AGES sample have companions within a projected distance of $30~{\rm kpc}$, we find that 
8 (33\%) of the 24 K+A galaxies have companion galaxies. The probability of randomly finding this
enhanced number of companion galaxies is only 4\%, so we conclude
that K+A galaxies have more companions than the bulk of the galaxy population.

We also visually inspected NDWFS images of K+A galaxies for additional evidence 
of galaxy mergers, including tidal tails,  and in Figure~\ref{fig:cutout} 
we provide $20^{\prime\prime}$ $I$-band postage stamp images of the K+A galaxies. 
Ten (42\%) of the K+A galaxies display tidal tails, including 7 that do not have companions within 
a projected distance of $30~{\rm kpc}$. The majority of K+A galaxies in our sample have 
nearby companion galaxies or tidal tails.

The high fraction of K+A galaxies undergoing mergers is broadly consistent with several previous studies,
although these studies use a diverse range of galaxy selection criteria.
\markcite{zab96}{Zabludoff} {et~al.} (1996) and \markcite{yang08}{Yang} {et~al.} (2008) find that a quarter 
and a half (respectively) of K+A galaxies show morphological evidence for ongoing mergers 
while \markcite{liu95}{Liu} \& {Kennicutt} (1995) found that a fifth of merging 
galaxies have K+A spectra. In contrast, \markcite{hog06}{Hogg} {et~al.} (2006) find that the environments of K+A galaxies
do not differ from those of star-forming galaxies, although the SDSS sample used by \markcite{hog06}{Hogg} {et~al.} (2006) 
is too shallow to detect many of the companion galaxies we find using the deeper NDWFS imaging.
While we find a higher fraction of merging K+A galaxies than prior studies, this is not unexpected
as many K+A samples are small, the depth of their imaging varies,  and the selection criteria for both
K+A galaxies and mergers differ.

The fraction of merging K+A galaxies appears to increase with luminosity, with 
all but one of the eight $M_R<-22$ K+A galaxies having a tidal tail or a cataloged
companion galaxy. However, the evidence is intriguing rather than conclusive.
If the fraction of K+A galaxies undergoing mergers is 62\% and does not vary with luminosity,
then the probability that 7 of the 8 most luminous K+A galaxies in a sample would be undergoing mergers is 12\%. 
Furthermore, some of the faintest K+A galaxies in our sample are undergoing mergers.
It thus remains plausible that the K+A galaxy merger rate does not vary with optical luminosity.

\begin{figure}
\plotone{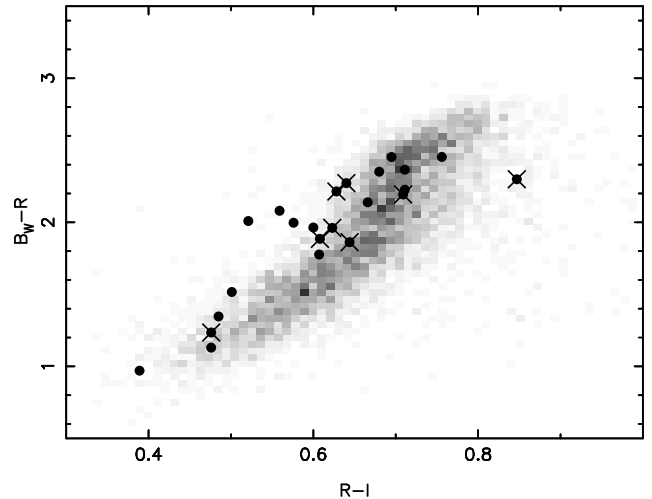}
\caption{The apparent colors of $0.10<z<0.35$ K+A galaxies (black circles), K+A galaxies with X-ray counterparts (crosses) 
and other AGES galaxies (greyscale). One K+A galaxy has been excluded from the plot as its photometry is 
contaminated by a nearby bright star. The K+A galaxies have a broader spread of $B_W-R$ and $R-I$ colors than other galaxies,
and it thus may be possible to select a large but incomplete sample of K+A candidates using apparent color criteria, 
although spectroscopy will still be required to produce a clean K+A sample. 
\label{fig:bwri}}
\end{figure}

We determined the apparent optical colors of our galaxies using copies of the NDWFS optical 
images smoothed to a common point spread function with  a FWHM of $1.35^{\prime\prime}$ and 
$8^{\prime\prime}$ aperture photometry \markcite{bro07}({Brown} {et~al.} 2007). 
In Figure~\ref{fig:bwri} we plot the apparent colors of our K+A galaxies and all 
$0.10<z<0.35$ AGES galaxies for comparison. 
We note that K+A galaxies have a broader spread of apparent colors than other AGES galaxies, which 
may be due to K+A galaxies having a prominent Balmer break rather than a strong 4000~\AA~ break.
Others have noted that the colors of post-starbursts can differ from those of normal galaxies, and the 
K+A galaxies of \markcite{tre07}{Tremonti} {et~al.} (2007) were initially targeted (as quasar candidates) by the Sloan 
Digital Sky Survey for this reason. One may be able to isolate a large (albeit incomplete) sample 
of K+A galaxy candidates on the basis of their apparent colors and extended morphologies, although 
spectroscopy will still be required to obtain a clean K+A sample. 

\begin{figure}
\plotone{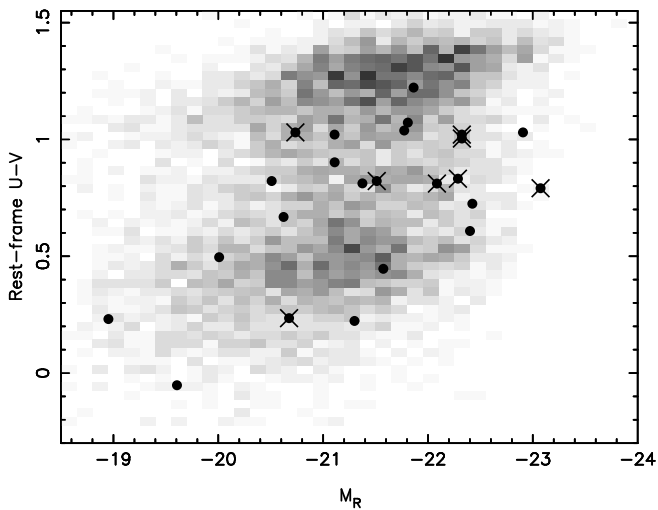}
\caption{Rest-frame $U-V$ color as a function of $R$-band absolute magnitude.
K+A galaxies (black circles) include some of the most and least luminous galaxies in the $0.10<z<0.35$ AGES sample.  
K+A galaxies with counterparts in the {\it Chandra} X-ray imaging (crosses) are, on average, more optically luminous than the 
bulk of the sample. As expected, K+A galaxies mostly fall between the red passive galaxies and the blue ``cloud'' of star forming galaxies,
although K+A galaxies are a small minority of the galaxies in this color range.
\label{fig:rest}}
\end{figure}

To determine the rest-frame photometry of our galaxies, we fitted \markcite{col80}{Coleman}, {Wu}, \& {Weedman} (1980) template spectra to the 
optical photometry. To span the locus of galaxy colors, we interpolated between 
the \markcite{col80}{Coleman} {et~al.} (1980) templates and extrapolated the templates to bluer and redder colors. We caution that 
the apparent colors of K+A galaxies do differ from those of most galaxies, so it is plausible
that our rest-frame photometry for K+A galaxies has small systematic errors. For example, 
the Balmer break in K+A galaxies may have been incorrectly modeled with a 4000~\AA~ break during the SED fitting. 
In Figure~\ref{fig:rest} we plot rest-frame $U-V$ as a function of $R$-band absolute magnitude, which is strongly
correlated with stellar mass. As seen in previous studies \markcite{qui04,yan08}(e.g., {Quintero} {et~al.} 2004; {Yan} {et~al.} 2008), most K+A galaxies have 
colors which fall between the ``blue cloud'' of star forming galaxies and the color-magnitude relation of red passive
galaxies. This is not unexpected, as we have selected galaxies with negligible rates of star formation with stellar
populations that are younger than those of red passive galaxies.

The stellar mass contained within the red galaxy population doubles between $z=1$ and $z=0$, due to 
steady transfer of stellar mass from the blue galaxy population to the red galaxy 
population \markcite{bel04,bro07,fab07}({Bell} {et~al.} 2004; {Brown} {et~al.} 2007; {Faber} {et~al.} 2007). 
Consequently, many of the galaxies with intermediate colors must be in transition between the blue and red populations.
We find that K+A galaxies represent just 1\% of the 1984 $0.10<z<0.35$ AGES galaxies with high signal-to-noise 
spectra and restframe colors in the range $0.6<U-V<1.1$. 
If all galaxies moving between the blue and red populations had their star formation abruptly truncated ($<300~ {\rm Myr}$), 
so they satisfied K+A selection criteria, and it takes $\sim 1~{\rm Gyr}$ for the $U-V$ colors of galaxies to evolve from blue
to red \markcite{bel04}(e.g., {Bell} {et~al.} 2004),  we should have identified hundreds of K+A galaxies.
Since we have only selected 24 K+A galaxies, we conclude that most of the galaxies falling between the
red and blue populations at $z\sim 0.2$ have had their star formation decline gradually (so they do not meet K+A selection criteria). 
This interpretation is consistent with the recent work of \markcite{yan08}{Yan} {et~al.} (2008), 
who find that the slowly evolving environments of post-starbursts differ from those of most red galaxies. 
In contrast to our findings, \markcite{wil09}{Wild} {et al.} (2009) find that K+A galaxies do play a significant 
role in transferring stellar mass from the blue population to the red population at $z\sim 0.7$, although this would 
require the space density of K+A galaxies to evolve by two orders of magnitude since $z=0.7$.
We conclude that while $z\sim 0.2$ K+A galaxies are in transition from the blue galaxy population
to the red galaxy population, this transition seems to be atypical and unusually rapid for low redshift galaxies.


\section{OPTICAL EMISSION-LINE FLUX RATIOS}
\label{sec:bpt}

The relative strength of emission-lines in a spectrum will vary depending upon the spectral energy 
distribution of the photo-ionizing radiation. As a consequence, emission line ratios can be used to 
distinguish AGNs from other galaxies \markcite{bal81,kew01,kau03}(e.g., {Baldwin} {et~al.} 1981; {Kewley} {et~al.} 2001; {Kauffmann} {et~al.} 2003), although there are caveats.
The classifications derived from emission-line ratios can vary with the size of the aperture used when 
obtaining the spectrum \markcite{zar95}(aperture bias; e.g., {Zaritsky}, {Zabludoff}, \&  {Willick} 1995), 
with weak AGNs being swamped by starlight in large aperture spectra. For most of the galaxies
in our sample, the Hectospec fiber is capturing 10\% to 30\% of the total galaxy light, so the variation
in aperture bias from galaxy-to-galaxy within our sample is small. Emission-line ratios also 
depend on metallicity  and the distributions of gas, dust and photoionizing sources.
Despite these limitations, emission-line ratios remain a useful diagnostic of AGN content.

In this work we use the BPT diagram \markcite{bal81}({Baldwin} {et~al.} 1981), which uses the emission-line flux ratios 
of ${\rm [NII]}~\lambda 6583$ divided by ${\rm H \alpha}$ and ${\rm [O~III]}~\lambda 5007$ divided by 
${\rm H\beta}$. Emission-line fluxes were determined by subtracting the stellar continuum and simultaneously 
fitting a Gaussian to each of the emission-line profiles \markcite{mou06}({Moustakas} \& {Kennicutt} 2006, J.~Moustakas, in prep.).
This technique results in emission-line measurements reliably corrected for stellar absorption.
The stellar continuum was modeled using a non-negative linear combination of high resolution \markcite{bc03}{Bruzual} \& {Charlot} (2003)
stellar population synthesis models (which span a wide range of stellar population ages), with dust attenuation being modeled using the
prescription outlined by  \markcite{mou06}{Moustakas} \& {Kennicutt} (2006). With the AGES spectra, we are able to determine accurate 
emission-line fluxes when the line strength is greater than $\simeq 5\times 10^{-17}~{\rm erg~s^{-1}~cm^{-2}}$.
We note that the strength of the  ${\rm H \alpha}$ and ${\rm H\beta}$ nebular emission-lines are sensitive to errors in the 
estimated strength of the Balmer absorption lines produced by the underlying stellar continuum model (\S\ref{sec:sample}). 
We also caution that an AGN can be missed if starlight and nebular emission dominates the light within the $1.5^{\prime\prime}$ 
aperture of the Hectospec fiber.

\begin{figure}
\plotone{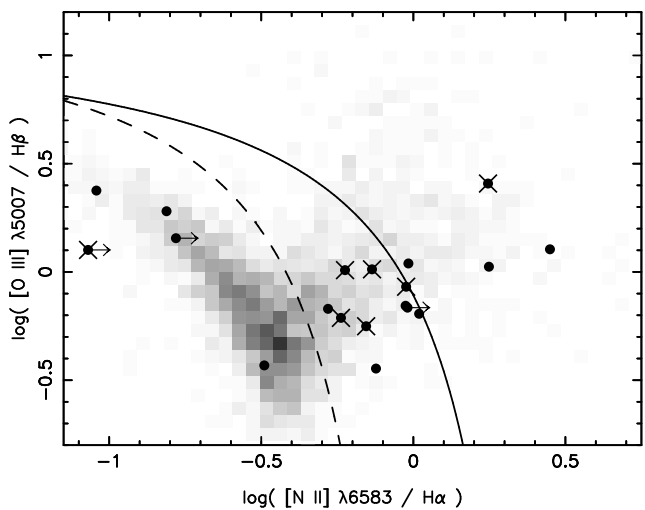}
\caption{BPT diagram of K+A galaxies (black circles), K+A galaxies with X-ray counterparts (crosses), and
other AGES galaxies (greyscale). For clarity we only plot those sources with $2\sigma$ emission-line detections or limits.
The solid line is a theoretical upper limit for what can be produced by a stellar 
population \markcite{kew01}({Kewley} {et~al.} 2001), while the dashed line is an empirical criterion used to distinguish AGNs from other galaxies
\markcite{kau03}({Kauffmann} {et~al.} 2003). The vast majority of our K+A galaxies lie in the AGN region of the BPT
diagram defined using the \markcite{kau03}{Kauffmann} {et~al.} (2003) criterion. 
\label{fig:bpt}}
\end{figure}

In Figure~\ref{fig:bpt} we plot the BPT diagram of $0.10<z<0.35$ K+A galaxies and other AGES galaxies. 
For comparison, we plot the upper limit for the emission-line ratios produced by gas photoionized by 
massive stars from \markcite{kew01}{Kewley} {et~al.} (2001) and the empirical AGN criterion of \markcite{kau03}{Kauffmann} {et~al.} (2003). Clearly the K+A galaxy population
differs significantly from the bulk of the galaxy population. Half of the K+A galaxies lie
beyond the stellar population limit of \markcite{kew01}{Kewley} {et~al.} (2001), and the vast majority lie in the AGN 
portion of the BPT diagram defined with the \markcite{kau03}{Kauffmann} {et~al.} (2003) criterion. Our BPT diagram is
broadly similar to that of \markcite{yan06}{Yan} {et~al.} (2006), except we have slightly fewer sources with strong
${\rm [O~III]}~\lambda 5007$ emission relative to ${\rm H\beta}$. While we are limited by small number statistics, it 
does appear that emission-line ratios may vary with optical luminosity. 
All three of the $M_R>-20.5$ K+A galaxies in our sample have ${\rm log ([NII]~\lambda~6583~/~H\alpha)}\sim -1$,
which is consistent with the nebular emission-line ratios produced by a stellar population.
While it is possible that the $M_R>-20.5$ K+A galaxies are harboring weak AGNs that are swamped
by starlight and associated nebular emission, or hosted powerful AGNs in the past, it 
is unlikely that these galaxies are currently hosting powerful AGNs.
If the emission-line ratios measured by AGES are a reliable diagnostic of AGN activity, the fraction of K+A galaxies
currently hosting an AGN increases with optical luminosity, with most $M_R<-20.5$ K+A galaxies hosting an AGN.

\section{CHANDRA X-RAY IMAGING}
\label{sec:chandra}

AGNs which are not evident at optical wavelengths may be detected with X-ray imaging \markcite{bar01}(e.g., {Barger} {et~al.} 2001).
The {\it Chandra X-ray Observatory} has imaged the entire Bo\"otes field, with the lion's share of the imaging
coming from the XBo\"otes wide-field and XBo\"otes deep surveys. We refer the reader to the XBo\"otes survey 
papers \markcite{ken05,mur05,bra06,mur06}({Kenter} {et~al.} 2005; {Murray} {et~al.} 2005; {Brand} {et~al.} 2006; {Murray} \& {XBootes Team} 2006) for a thorough discussion of the observing strategy and data processing.
The exposure time per pointing is at least $4~{\rm ks}$, with some regions of Bo\"otes having 
exposure times as high as $172~{\rm ks}$ \markcite{wan04}({Wang} {et~al.} 2004). 
None of the K+A galaxies are associated with known extended X-ray sources in the Bo\"otes field \markcite{ken05}({Kenter} {et~al.} 2005).
Eight (33\%) of the K+A galaxies have one or more X-ray photons within $2^{\prime\prime}$ of the optical position, 
including five (17\%) K+A galaxies with two or more X-ray photons. For comparison, 995 of the 6592 (15\%) AGES galaxies 
have one or more X-ray photons within $2^{\prime\prime}$ of the optical position, with 316 (5\%) having two or 
more X-ray photons. K+A galaxies are thus more likely to harbor X-ray sources than the overall galaxy population.

We determined fraction of K+A galaxies with spurious X-ray counterparts by searching for X-ray photons
within $2^{\prime\prime}$ of positions offset by $10^{\prime\prime}$ from the K+A galaxies. Approximately 8\%
of the offset positions had spurious X-ray counterparts (usually single photon detections) and we thus
conclude that $\simeq 2$ of the 24 K+A galaxies also have spurious X-ray counterparts.
We note that the only X-ray source that does not lie in the AGN portion of the BPT diagram defined 
with the \markcite{kau03}{Kauffmann} {et~al.} (2003) criterion (see Figure \ref{fig:bpt}) 
is a one photon source, and is thus likely to be spurious. This is also the only X-ray source that does not show
evidence for an ongoing galaxy merger in the optical imaging. 

\begin{figure}
\plotone{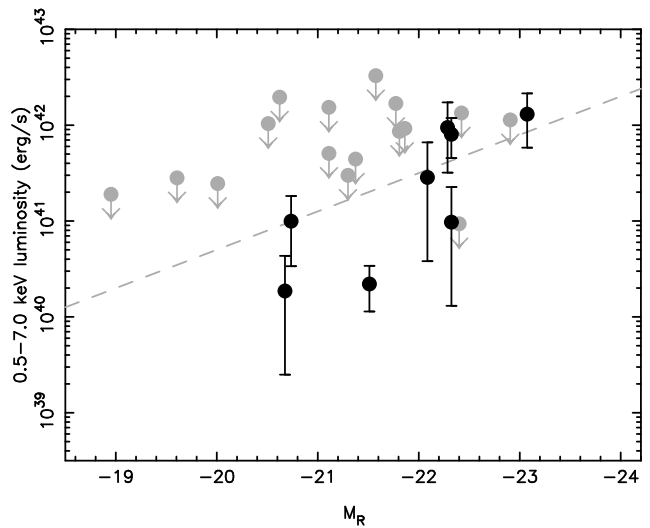}
\caption{The X-ray luminosities of K+A galaxies. The {\it Chandra} exposure times vary by an order 
of magnitude, so our uncertainties and $2\sigma$ upper limits vary accordingly. As we are using 
a flux limited sample, there is a correlation between absolute magnitude and 
limiting X-ray luminosity. We expect $\sim 2$ spurious X-ray counterparts in the 
sample, and that these sources will be single photon detections.
The dashed line denotes the trend expected if X-ray luminosity divided by
optical luminosity is constant. Five of the eight K+A galaxies with absolute magnitudes
brighter than $M_R=-22$ have X-ray counterparts, and at least three have X-ray luminosities
on the order of $10^{42}~{\rm erg~s^{-1}}$. These powerful X-ray sources are almost certainly AGNs.
\label{fig:xlum}}
\end{figure}

To determine the nature of the X-ray sources, we estimated their X-ray fluxes and luminosities (or $2\sigma$ upper limits).
To constrain the photon index, we determined the hardness ratio,
\begin{equation}
H.R. = (H-S)/(H+S),
\end{equation}
where $S$ and $H$ are the photon counts in the 0.5-2.0~keV and 2.0-7.0~keV bands respectively. Using the 19 photons 
from the 8 K+A galaxies, we measure a hardness ratio of $-0.2\pm 0.5$, corresponding to a photon index 
of $\Gamma=1.0\pm 1.0$. We used PIMMS version 3.9 \markcite{muk93}({Mukai} 1993) to determine the 
conversion from ({\it Chandra} Cycle~4) photon counts to unabsorbed X-ray flux, using a source 
photon index of $\Gamma=1.0$ and a Galactic foreground HI column density of $1.75\times 10^{20}~{\rm cm^{-2}}$ \markcite{sta92}({Stark} {et~al.} 1992). 
X-ray luminosities were determined using
\begin{equation}
L_X = 4\pi d_L^2 f_X (1+z)^{\Gamma-2},
\end{equation}
where $d_L$ is the luminosity distance and $f_X$ is the X-ray flux. Although our estimate of the photon index is highly uncertain,
this has a modest impact on our luminosities as we are studying low redshift objects and our photon shot noise is large.
The X-ray luminosities of our sources are plotted in Figure~\ref{fig:xlum} and listed in Table~\ref{table:summary}. 

We find that the X-ray luminosities of K+A galaxies are, on average, higher than those of other galaxies.
The mean X-ray luminosity of K+A galaxies (determined using galaxies with and without XBo\"otes detections) 
is $1.5\times 10^{41}~{\rm erg~s^{-1}}$  while the mean X-ray luminosity of other $I<20$ AGES galaxies is 
$7\times 10^{40}~{\rm erg~s^{-1}}$. (We caution that individual galaxies can have luminosities far higher and lower
than these mean values.) The mean X-ray luminosity of K+A galaxies brighter than $M_R=-22$ is $4\times 10^{41} {\rm erg~s^{-1}}$, while 
other AGES galaxies with comparable absolute magnitudes have a mean X-ray luminosity of $1.9\times 10^{41} {\rm erg~s^{-1}}$.
This is broadly consistent with the recent stacking analysis of \markcite{geo08}{Georgakakis} {et~al.} (2008), who find that the mean X-ray 
luminosity of $z\sim 0.8$ post-starbursts is $\sim 3\times 10^{41}~{\rm erg~s^{-1}}$. 

Many of the K+A galaxies with X-ray counterparts are almost certainly hosting AGNs.
Several K+A galaxies have X-ray luminosities of $10^{42}~{\rm erg~s^{-1}}$, which is a factor
of $\sim 4$ higher than local ultraluminous infrared galaxies 
that are predominantly powered by ongoing star formation \markcite{pta03}({Ptak} {et~al.} 2003).
Most of the K+A galaxies with X-ray counterparts would also be classified as AGNs using 
the BPT diagram and the empirical criterion of \markcite{kau03}{Kauffmann} {et~al.} (2003).
As noted in \S\ref{sec:sample}, additional K+A galaxies hosting powerful AGNs are missing 
from our sample, due to our $H\alpha$ selection criterion (e.g., IRAS~14344+3451).

The X-ray luminosities of K+A galaxies are correlated with their optical luminosities.
Five of the eight K+A galaxies brighter than $M_R=-22$ have counterparts in the {\it Chandra} imaging,
with three having X-ray luminosities on the order of $10^{42}~{\rm erg~s^{-1}}$. 
K+A galaxies that are fainter than $M_R=-22$ have X-ray luminosities between $10^{40}~{\rm erg~s^{-1}}$ 
and $10^{41}~{\rm erg~s^{-1}}$, or have upper limits consistent with this luminosity range. 
The X-ray luminosities of K+A galaxies could be proportional to their optical luminosities, 
although steeper relationships are also consistent with our observations. 

The most luminous K+A galaxies are frequently undergoing mergers and show strong evidence for hosting AGNs,
including powerful X-ray emission. Many of the faintest K+A galaxies do not show evidence for ongoing
mergers nor AGN activity. Perhaps AGN activity is weak or brief in the faintest K+A galaxies, so it
is difficult to observe. Alternatively, we can conclude that the fraction of K+A galaxies
that have hosted powerful AGNs and undergone mergers does increase with optical luminosity.
If this is the case, perhaps two mechanisms are responsible for the truncation of star formation, with 
one being correlated with galaxy mergers and AGN activity. \markcite{yan08}{Yan} {et~al.} (2008) also conclude that there may be two mechanisms for producing
K+A galaxies, after noting a bimodality in the distribution of K+A environments. Multiple mechanisms for
the truncation of star formation are also found in recent semi-analytic models \markcite{bow06,cro06}(e.g., {Bower} {et~al.} 2006; {Croton} {et~al.} 2006),
with AGN feedback only playing a major role in central galaxies while other mechanisms regulate 
star formation in satellite and low luminosity galaxies, although how accurately these models mimic
galaxy evolution remains a matter of vigorous debate.

\section{SUMMARY}
\label{sec:summary}

We have examined the AGN content of K+A galaxies, using 24 K+A galaxies 
selected from AGES spectroscopy of $7.9~{\rm deg}^2$ of the Bo\"otes field.
Using deep optical imaging from the NDWFS, we find that two thirds of 
K+A galaxies are likely ongoing mergers or are recent merger remnants. 
As the merger of two galaxies (initially separated by $\sim 30~{\rm kpc}$) 
can take as long as a billion years, 
it is plausible that the mechanism that truncates star formation is triggered by 
galaxy mergers or that we are observing the aftermath of merger triggered bursts of star formation.

Between $z=1$ and $z=0$ the stellar mass contained within the red galaxy population
doubles, due to the transfer of stellar mass from the blue star forming galaxy population to the
red passive galaxy population. As the star formation rates of galaxies decline and they 
move towards the red galaxy population, their rest-frame optical colors will fall between the
red and blue populations. While the optical colors of K+A galaxies fall between the red and 
blue populations, they represent just 1\% of the $z\sim 0.2$ galaxies with these optical colors.
We thus conclude that, compared to most galaxies moving from the blue population 
to the red population, the truncation of star formation in K+A galaxies may be unusually abrupt.

We find that K+A galaxies frequently host AGNs. Half of the K+A galaxies in 
our sample have emission-line ratios comparable to those of Seyferts and LINERs, while a 
quarter have counterparts in the {\it Chandra} XBo\"otes survey.
While not direct evidence for AGN feedback, it is consistent with AGN feedback being responsible 
for the truncation or regulation of star formation in some galaxies.

The fraction of K+A galaxies that show clear evidence of hosting an AGN increases with optical luminosity.
While most K+A galaxies have optical emission-line ratios consistent with AGNs, the faintest K+A galaxies
have emission-line ratios consistent with those produced by a stellar population.
A third of $M_R<-22$ K+A galaxies have X-ray counterparts with luminosities of 
$\sim 10^{42}~{\rm erg~s^{-1}}$, while it is rare for optically faint K+A galaxies to have X-ray counterparts.
It is possible that the faintest K+A galaxies host short lived or very faint AGNs, which 
are not detected by XBo\"otes and are swamped by starlight and nebular emission in optical spectra.
Alternatively, we speculate that two mechanisms may be responsible for truncating star formation in K+A galaxies, 
with AGNs truncating star formation in the most luminous galaxies while another mechanism truncates star formation 
in lower luminosity galaxies.

\acknowledgments

We thank our colleagues on the AGES, NDWFS, and XBo\"otes teams. This paper would not have been possible 
without the efforts of the  {\it Chandra}, KPNO, and MMT support staff.
David Palamara's involvement in this paper began during his studies for Monash University's PHS3360 unit, which was coordinated by John Cashion.
Late in the development of this work, Richard Cool was supported by NASA through Hubble Fellowship grant HST-HF-01217.01,
awarded by the Space Telescope Science Institute, which is operated by the Association of Universities for Research in 
Astronomy, for NASA, under the contract NAS 5-26555.
John Moustakas received support from NASA grant 06-GALEX06-0030 and Spitzer grant G05-AR-50443 during the development of this paper.
Spectroscopic observations reported here were obtained at the MMT Observatory, a joint facility of the Smithsonian Institution and the University of Arizona.
This work made use of optical images provided by the NOAO Deep Wide-Field Survey \markcite{jan99}({Jannuzi} \& {Dey} 1999), which is supported by the National Optical 
Astronomy Observatory (NOAO). NOAO is operated by AURA, Inc., under a cooperative agreement with the National Science Foundation.


\bibliography{}

\begin{deluxetable}{cccccccccc}
\tablecolumns{10}
\tabletypesize{\tiny}
\tablecaption{Bo\"otes K+A galaxy sample, ordered by absolute magnitude.\label{table:summary}}
\tablehead{
  \colhead{J2000 Coordinates}      &
  \colhead{Redshift}               &
  \colhead{$I$}                    &
  \colhead{$M_R$}                  &
  \colhead{Companion}              &
  \colhead{Tidal}             &
  \colhead{BPT\tablenotemark{b}}             &
  \colhead{X-ray}          &
  \colhead{X-ray Exp.}             & 
  \colhead{$L_X$} \\
  \colhead{}               &
  \colhead{}               &
  \colhead{}               &
  \colhead{}               &
  \colhead{($<30~{\rm kpc}$)} &
  \colhead{Tail}               &
  \colhead{Class}              &
  \colhead{Photons}           &
  \colhead{$({\rm sec.})$}    & 
  \colhead{$({\rm erg~s^{-1}}$} 
}
\startdata
14:26:37.266 +34:46:25.63  & 0.1314 & 19.45 &  -18.95 &  N &  N &  SF     &  0 &  8943 &  $ < 1.9 \times 10^{41} $\\ 
14:26:18.398 +35:02:48.55  & 0.1567 & 19.31 &  -19.61 &  Y &  N &  SF     &  0 &  8658 &  $ < 2.8 \times 10^{41} $\\ 
14:30:14.018 +33:51:38.84  & 0.1817 & 19.01\tablenotemark{a} &  -20.01 &  Y &  N &  SF     &  0 & 13481 &  $ < 2.5 \times 10^{41} $\\ 
14:35:51.378 +33:33:53.72  & 0.2037 & 18.93 &  -20.51 &  N &  N &  ?      &  0 &  4055 &  $ < 1.0 \times 10^{42} $\\ 
14:38:11.278 +34:07:47.69  & 0.2641 & 19.52 &  -20.62 &  N &  N &  ?      &  0 &  3694 &  $ < 2.0 \times 10^{42} $\\ 
14:33:29.167 +33:28:21.37  & 0.1725 & 18.36 &  -20.68 &  N &  N &  SF     &  1 & 42306 &  $ (1.9 \pm^{2.5}_{1.6}) \times 10^{40} $\\ 
14:29:47.738 +34:16:18.81  & 0.1220 & 17.41 &  -20.74 &  N &  Y &  SF/AGN &  2 &  7722 &  $ (1.0 \pm^{0.8}_{0.7}) \times 10^{41} $\\ 
14:38:19.445 +35:36:01.98  & 0.2483 & 18.88 &  -21.11 &  N &  Y &  SF/AGN &  0 &  4141 &  $ < 1.5 \times 10^{42} $\\ 
14:30:55.228 +33:26:01.11  & 0.2017 & 18.35 &  -21.11 &  N &  Y &  ?      &  0 &  8126 &  $ < 5.1 \times 10^{41} $\\ 
14:31:01.102 +34:08:44.34  & 0.2307 & 18.47 &  -21.30 &  N &  N &  SF     &  0 & 18232 &  $ < 3.0 \times 10^{41} $\\ 
14:25:52.846 +32:55:56.05  & 0.2139 & 18.20 &  -21.37 &  N &  N &  ?      &  0 & 10536 &  $ < 4.4 \times 10^{41} $\\ 
14:25:48.378 +35:40:55.65  & 0.1840 & 17.66 &  -21.51 &  N &  Y &  SF/AGN &  4 & 163416 &  $ (2.2 \pm^{1.2}_{1.1}) \times 10^{40} $\\ 
14:34:39.522 +34:30:21.67  & 0.3411 & 19.19 &  -21.57 &  N &  N &  SF/AGN &  0 &  3773 &  $ < 3.3 \times 10^{42} $\\ 
14:31:20.690 +35:29:26.20  & 0.3347 & 19.00 &  -21.77 &  Y &  N &  SF/AGN &  0 &  7068 &  $ < 1.7 \times 10^{42} $\\ 
14:29:59.370 +34:22:41.67  & 0.2655 & 18.41 &  -21.81 &  N &  N &  AGN    &  0 &  8511 &  $ < 8.6 \times 10^{41} $\\ 
14:35:00.640 +33:29:23.18  & 0.2736 & 18.33 &  -21.86 &  Y &  N &  SF/AGN &  0 &  8430 &  $ < 9.3 \times 10^{41} $\\ 
14:25:42.236 +32:44:08.47  & 0.2079 & 17.47 &  -22.09 &  N &  Y &  SF/AGN &  1 &  4078 &  $ (2.8 \pm^{3.8}_{2.5}) \times 10^{41} $\\ 
14:34:54.275 +33:00:59.97  & 0.2986 & 18.22 &  -22.29 &  N &  N &  ?      &  2 &  5239 &  $ (9.5 \pm^{7.8}_{6.3}) \times 10^{41} $\\ 
14:25:39.281 +33:56:44.48  & 0.3085 & 18.20 &  -22.32 &  Y &  Y &  SF/AGN &  5 & 16506 &  $ (8.0 \pm^{3.9}_{3.5}) \times 10^{41} $\\ 
14:29:26.729 +33:48:30.14  & 0.2417 & 17.56 &  -22.32 &  Y &  Y &  SF/AGN &  1 & 16328 &  $ (9.7 \pm^{12.9}_{8.4}) \times 10^{40} $\\ 
14:31:46.450 +34:17:18.49  & 0.1234 & 15.80 &  -22.40 &  N &  Y &  AGN    &  0 & 15953 &  $ < 9.4 \times 10^{40} $\\ 
14:34:31.800 +33:47:59.17  & 0.3318 & 18.31 &  -22.42 &  Y &  N &  SF/AGN &  0 &  8725 &  $ < 1.3 \times 10^{42} $\\ 
14:25:44.882 +33:34:31.20  & 0.3011 & 17.59 &  -22.91 &  Y &  Y &  AGN    &  0 &  8391 &  $ < 1.1 \times 10^{42} $\\ 
14:26:21.757 +35:11:47.37  & 0.3461 & 17.85 &  -23.07 &  N &  Y &  AGN    &  3 &  7772 &  $ (1.3 \pm^{0.8}_{0.7}) \times 10^{42} $\\ 
\hline
\enddata
\tablenotetext{a}{A nearby bright star may be biasing the photometry of this galaxy.}
\tablenotetext{b}{AGN denotes objects above the criterion of \markcite{kew01}{Kewley} {et~al.} (2001). SF/AGN denoted objects between the criteria of  \markcite{kew01}({Kewley} {et~al.} 2001) and \markcite{kau03}({Kauffmann} {et~al.} 2003). SF denotes objects below the criterion of \markcite{kau03}{Kauffmann} {et~al.} (2003). ``?'' denotes objects without significant detections in the relevant emission-lines for the BPT diagram.}
\end{deluxetable}

\clearpage

\end{document}